\documentclass[a4paper,fleqn,usenatbib]{mnras}

\usepackage{newtxtext,newtxmath}
\usepackage{multirow}
\usepackage[T1]{fontenc}
\usepackage{ae,aecompl}

\usepackage{graphicx}	% Including figure files
\usepackage{amsmath}	% Advanced maths commands
  \title[Properties of IOAG candidates]{AGN and Star-Formation Properties of Inside-out Assembled Galaxy Candidates at z\,$<$\,0.1} 

\author[Zewdie et al.]{
Dejene Zewdie$^{1,2,3}$\thanks{E-mail: dejene.woldeyes@mail.udp.cl},
Mirjana Povi\'c$^{3,4}$,
Manuel Aravena$^{1}$,
Roberto J. Assef$^{1}$ and
\newauthor Asrate Gaulle$^{5}$
\\
$^{1}$ N\'ucleo de Astronom\'ia de la Facultad de Ingenier\'ia y Ciencias, Universidad Diego Portales, Av. Ej\'ercito Libertador 441, Santiago, Chile\\
$^{2}$Department of Physics, College of Natural and Computational Science, Debre Berhan University (DBU), Debre Berhan, Ethiopia\\
$^{3}$Astronomy and Astrophysics Research and Development Division, Entoto Observatory and Research Center (EORC),\\  \hspace{0.15cm} Ethiopian Space Science and Technology Institute (ESSTI), Algeria St., P.O.Box 33679, Addis Ababa, Ethiopia\\
$^{4}$ Instituto de Astrof\'isica de Andaluc\'ia (IAA-CSIC), Glorieta de la Astronom\'ia s/n, 18008, Granada, Spain\\
$^{5}$  Department of Physics, College of Natural and Computational Science, Dilla University, Dilla, Ethiopia
}

\date{Accepted 2020 August 12. Received 2020 August 12; in original form 2019 November 25}

\pubyear{2020}

\begin{document}
\label{firstpage}
\pagerange{\pageref{firstpage}--\pageref{lastpage}}
\maketitle

% Abstract of the paper
\begin{abstract}

We study a sample of 48127 galaxies selected from the SDSS MPA-JHU catalogue, with $\log M_{\star}/M_{\odot} = 10.73 - 11.03$ and $z<0.1$. Local galaxies in this stellar mass range have been shown to have systematically shorter assembly times within their inner regions ($<0.5~R_{50}$) when compared to that of the galaxy as a whole, contrary to lower or higher mass galaxies which show consistent assembly times at all radii. Hence, we refer to these galaxies as Inside-Out Assembled Galaxy (IOAG) candidates. We find that the majority of IOAG candidates with well-detected emission lines are classified as either AGN (40\%) or composite (40\%) in the BPT diagram. We also find that the majority of our sources are located below the main sequence of star formation, and within the green valley or red sequence. Most BPT-classified star-forming IOAG candidates have spiral morphologies and are in the main sequence, whereas Seyfert 2 and composites have mostly spiral morphologies but quiescent star formation rates (SFRs). We argue that a high fraction of IOAG candidates seem to be in the process of quenching, moving from the blue cloud to the red sequence. Those classified as AGN have systematically lower SFRs than star-forming galaxies suggesting that AGN activity may be related to this quenching. However, the spiral morphology of these galaxies remains in place, suggesting that the central star-formation is suppressed before the morphological transformation occurs.

\end{abstract}

\begin{keywords}
Galaxies: Properties -- Galaxies: Evolution -- Galaxies: Star formation
\end{keywords}

\section{Introduction} \label{s1}
How galaxies form and evolve through cosmic time is one of the major open questions in extragalactic astronomy and modern cosmology. In particular, we need to understand what role do galaxy mergers, AGN activity and star-formation feedback play in driving the observed morphological evolution and the quenching of star formation in massive galaxies. 

Over the last few decades, a number of important empirical correlations of galaxy parameters have been discovered. This includes the colour-stellar mass relation, in which galaxies show a bi-modal distribution (e.g., \citealt{2013povic}; \citealt{2014schawinski}; \citealt{2017Mahoro}, \citealt{2018Nogueira}). On one hand, late-type galaxies populate a region called the blue cloud, found at, as the name says, blue UV/optical colours and stellar masses typically below $\log \rm{M_{\star}}$ \,=\, 11 $\rm{M_{\odot}}$. On the other hand, early-type galaxies form a tight relation between their red UV/optical colours and their stellar masses, usually called the red sequence.  Between the blue cloud and red sequence there is an under-occupied space known as the ``green valley'', which includes a mixed population of galaxies (e.g., \citealt{Brinchmann};  \citealt{2012Whitaker}; \citealt{2013Guo}; \citealt{2014schawinski}; \citealt{2015Salim};   \citealt{2017Mahoro, 2019Mahoro}), most notably, post-starbursts galaxies have strong Balmer absorption lines  and the lack of emission lines (E+A galaxies, e.g., \citealt{1983Dressler}; \citealt{1996Zabludoff}; \citealt{Matsubayashi_2011}).  This bi-modality has been observed at both low and higher redshifts, up to at least z $\sim$ 2 (e.g., \citealt{Ilbert2010}; \citealt{2014schawinski}; \citealt{2017Mahoro}).

A tight relation between the star formation rates and stellar masses of typical star-forming galaxies has also been identified (e.g., \citealt{Brinchmann}; \citealt{2007Noeske}, \citealt{2007Elbaz}, \citealt{2007Daddi}; \citealt{2011Karim}; \citealt{2012Whitaker, 2014Whitaker};  \citealt{2015Schreiber}; \citealt{2016Leslie}; \citealt{2016povic}), usually referred to as the ``main-sequence'' (MS) of star-forming galaxies.  Galaxies above this sequence (i.e., with higher SFRs for their stellar mass) are typically called starbursts and those below (i.e., with lower SFRs for their stellar mass) are usually called passive or quiescent galaxies (e.g., \citealt{2012Gon}; \citealt{2013Moustakas}; \citealt{2016Leslie}). Galaxies that have halted their star-formation activity are usually called quenched galaxies. These galaxies can be broadly associated with those falling within the range of ``red passive galaxies'' or galaxies lying below the MS of star formation ($<$ 0.3 dex MS).

These relations suggest that the stellar mass is a fundamental parameter in the evolution of galaxies and that galaxies with different stellar masses may have very different formation and evolution histories. The evolution of the so called galaxy size - mass relation has been taken as evidence for inside out growth  (e.g., \citealt{Wel2014}; \citealt{Mowla2019}): at a fixed stellar mass, galaxies at high redshift seem to have been more compact than in the local universe. In addition, recent observations based on integral field spectroscopy and using the fossil record method (e.g., \citealt{2013Perez}; \citealt{Pan2015}; \citealt{IbarraMedel2016}; \citealt{GarcBenito2017}; \citealt{Liu2018};  \citealt{2018Sanchez}; \citealt{Wang2018}), have helped to characterise the mass assembly modes and star formation histories of galaxies as a function of radius. Evidence for segregated growth between the central and outer regions of massive galaxies (both on and off the main-sequence) has been found at high redshift (e.g., \citealt{IbarraMedel2016}; \citealt{Morselli2016}; \citealt{Nelson2016}; \citealt{Belfiore2018}; \citealt{Tacchella2018}) and as well as in simulations (e.g., \citealt{Aumer2014}; \citealt{AvilaReese2018}). \citet{2013Perez} used the integral field spectroscopic data from the Calar-Alto Legacy Integral Field Area (CALIFA) survey to study the growth rates of a sample of 105 galaxies in a range of stellar masses ($\log M_{\star}  = 9.58 - 11.26$ $ \rm{M_{\odot}}$) within different regions of the galaxies (nucleus, inner $0.5R_{50}$ and $R_{50}$\footnote{$R_{50}$ is the circular half-light radius in 5635 $\pm$ 40 {\AA}  \citep{2013Perez}.}, and outer region $>R_{50}$). \citet{2013Perez} found that in the stellar mass range $\log M_{\star}=10.73- 11.03$ $ \rm{M_{\odot} }$ the inner regions of galaxies reached 80\% of their final stellar mass twice as fast as in the outskirts.  They suggested different mechanisms to explain this growth, through the accretion of halo or intergalactic gas clouds, or through interactions and mergers with smaller or similar mass galaxies (see \citealt{2013Perez}). The authors suggested that perhaps more massive galaxies, that show inside-out growth, grow through minor and major mergers while in lower mass galaxies, where outside-in growth was observed, secular evolution could be the dominant mechanism. For the rest of this study, we will refer to the galaxies in this stellar mass range as ``Inside-Out Assembled Galaxy" (IOAG) candidates. In this work we suggest that these galaxies could play an important role in understanding their morphological transformation from late- to early-types.

Other studies suggest similar mass ranges of $\log M_{\star}  =  10.6-10.7$ $\rm{M_\odot}$ to be important in galaxy evolution, e.g., in the stellar-to-halo mass relation and accumulated stellar growth  (e.g., \citealt{ Mandelbaum2006}; \citealt{ Conroy2009}; \citealt{Moster2010}; \citealt{Behroozi2010};  \citealt{Guo2010}; \citealt{More2011}; \citealt{2012Leauthaud}), in simulations when studying the shape of stellar-to-halo mass relation and supernova (SN) and AGN feedback mechanisms  \citep{2006Shankar}, or when analysing low-ionization nuclear emission-line regions (LINERs) with high SFRs \citep{2016povic}.

In this work, we aim to characterise the physical properties of IOAG candidates and to understand better the nature of inside-out growth and its relation  with AGN activity.  This paper is organized as follows: In Section 2, we present the data, catalogues, and sample used throughout this work. In Section 3, we analyse the spectral classification of galaxies using optical emission lines, their morphological classification, SFR distributions, and their location in the SFR-stellar mass diagram as well as in the colour-stellar mass diagram. Our main results are discussed in Section 4, while our summary and conclusions are described in Section 5. Throughout this paper we assume a standard $\Lambda$ CDM cosmology with $H_0 = 70$ $\rm{km}$ $\rm{s^{-1}}$ $\rm{pc^{-1}}$, $\Omega_{\Lambda} =0.7$ and $\Omega_{\rm{M}} =0.3$, and we present all magnitudes in the AB system and the WISE Vega-system magnitudes.
 
\section{Data and sample selection} 
\subsection{SDSS}
In this work, we use data from the Sloan Digital Sky Survey (SDSS), Data Release (DR8; \citealt{2011ApJS}). The SDSS DR8 includes imaging in the `u', `g', `r', `i' and `z' photometric bands of 14,555 deg$^2$ throughout the sky, with roughly 5200 deg$^2$ in the southern Galactic cap and the rest in the northern hemisphere (\citealt{2000York}; \citealt{2011ApJS}). 

We use SDSS DR8 instead of a more recent Data Release due to the availability of estimates for a number of physical parameters through the MPA-JHU catalogue\footnote{http://www.sdss3.org/dr8/spectro/galspec.php}. Specifically, we use the emission line intensities, stellar masses, and SFRs provided through this catalogue. Stellar masses were calculated by \citet{2003Kauffmann} from the SDSS broad-band photometry and using a Bayesian Spectral Energy Distribution (SED) fitting methodology. We chose the stellar mass estimate corresponding to the median of the SED fit probability distribution function. The SFRs are calculated using a number of emission lines and the methods developed by \citet{Brinchmann}. For AGNs and galaxies with weak emission lines, the SFRs are computed using the correlation with Dn 4000 index (\citealt{2003Kauffmann}; \citealt{Brinchmann}) obtained through fiber aperture measurements and corrections based on broad band photometry \citep{2007salim}.

 \begin{figure}
\includegraphics [scale=0.44]{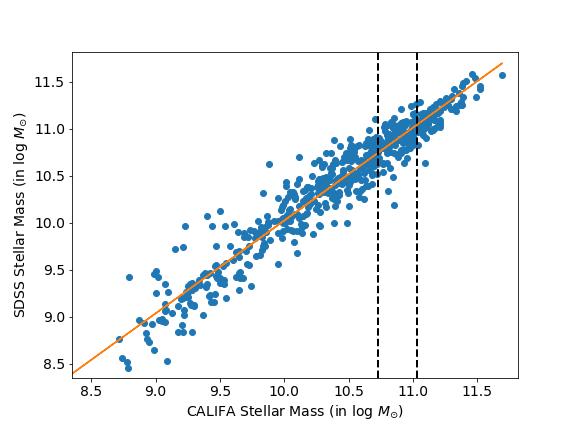}
\caption{Comparison between the stellar masses measured in the CALIFA survey \citep{2013Perez} and the SDSS MPA-JHU estimates \citep{Brinchmann} for a sample of 106 overlapping sources. The red solid line shows the 1-to-1 relation and the vertical dashed (black) lines the mass range of this study, based on the work of  \citet{2013Perez}. Given the low scatter between these estimates, and absence of obvious systematic offsets, the galaxy stellar masses used in our work can thus be validated.}
\label{fig1}
\end{figure}

\citet{2013Perez} results show the direct relation of relative assembly rate with stellar mass and suggest the highest differences in the proposed stellar-mass range of  $\log M*=10.73-11.03$ $M_{\odot}$, where the growth of the inner part of the galaxy is twice as faster as of the outer region. We noticed that narrower stellar mass range of log $\log M*=10.73-10.93$ $M_{\odot}$  with the relative assembly rate of $\sim 2.37$ could also work, however our selection of a bit larger stellar mass range was motivated by having significantly larger sample of galaxies to perform all statistical analysis. We tested all statistics for smaller-mass range as well, without finding any significant differences (not larger than 5\,-\,6\%), confirming that our results will remain consistent with the final selected range of stellar-mass. 

As mentioned earlier, this research is motivated by the results obtained by \citet{2013Perez}. To select IOAG candidates, we thus selected sources with stellar masses in the range $10.73 - 11.03$ $M_{\odot}$. The stellar mass range defined by \cite{2013Perez} was based on CALIFA data and used a different methodology than those used in the SDSS MPA-JHU stellar mass estimates. To check for possible biases in the stellar mass selection, we compare in Figure \ref{fig1} the \citet{2013Perez} stellar mass estimates with the MPA-JHU estimates for a sample of galaxies in the CALIFA survey. Both estimates are found to be in good agreement, with a scatter of $\sim$ 0.1 dex, thus validating the SDSS stellar mass measurements for this study and the selected mass range for the IOAG candidates.  

The photometric and spectroscopic SDSS galaxy catalogues were cross-matched using a radius of 2$''$ as the best compromise between the number of lost matches and possible miss-matched sources. We further restrict the sample to the IOAG candidates stellar mass range $10.73-11.03\ M_\odot$ and to $z<0.1$ leading to a final sample of 48127 galaxies. The upper limit in redshift was selected following Kewley et al. (2006) to avoid completeness issues with the spectroscopic classification.  The selected redshift range is also the proper one for our study when dealing with morphology (see Section \ref{morp}), knowing that the classification becomes challenging at higher redshifts, especially for imaging data from a shallow survey as SDSS \citep{2015Povic}.

\subsection{Spectral classification}\label{BPTsec}
 We classified galaxies spectroscopically using the ``Baldwin, Phillips \& Terlevich" (BPT) diagram \citep{1981Baldwin}. This classification is based on nebular emission-line ratios that are used to differentiate their ionizing source. To provide a clean BPT spectral classification and following previous studies, we applied a further restriction to our catalogue, by requiring that the signal-to-noise ratio (S/N) is above 3.0 for each of the emission lines used (e.g., \citealt{2001Kewley, 2006Kewley}; \citealt{Kauffmann2004}; \citealt{Brinchmann}). This results in samples of 24561, 18478, 11006 galaxies, respectively, for the [OIII]/H${\beta}$ versus [NII]/H${\alpha}$ (BPT-NII), [OIII]/H${\beta}$ versus [SII]/H${\alpha}$ (BPT-SII), and [OIII]/H${\beta}$ versus [OI]/H${\alpha}$ (BPT-OI) diagrams. Based on these line ratios, sources are classified as either star-forming, AGN (Seyfert 2 or LINER) or composite. Given the significantly larger number of sources in the BPT-NII diagram, we decided to focus on this classification for the rest of this study.  
 
 \begin{figure}
\includegraphics[scale=0.43]{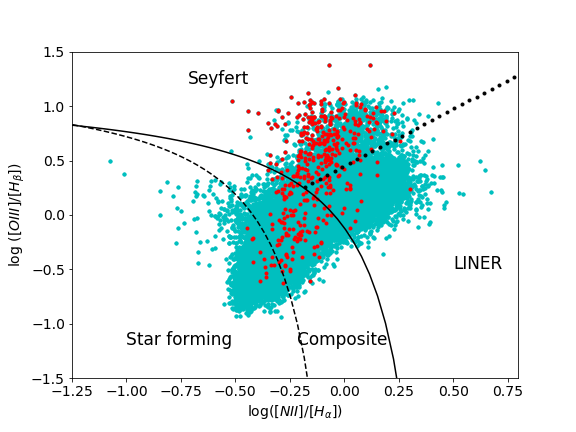}
\caption{The $[OIII]/H_{\beta}$ versus $[NII]/H_{\alpha}$BPT diagnostic diagram for the SDSS galaxies in our sample. The sample has been restricted to galaxies with line emission detected at the S/N $\geq 3$ level. The dashed line shows the \citep{2003Kauffmann} empirical division between star-forming and starburst/AGNs composite galaxies. The solid line represents the \citep{2006Kewley} division between star-forming galaxies and those with dominant AGNs contribution (extreme starburst limit). The dotted line shows the separation between Seyfert 2 and LINER \citep{2007Schawinski}. The cyan symbols represent the location of the selected SDSS galaxies in the stellar mass range occupied by inside-out assembled galaxies. The red symbols represent those galaxies that are WISE-selected AGN. As expected, the locus of WISE AGNs (413 \%).coincides with the Seyfert 2 AGNs optical classification. }
\label{fig2}
\end{figure}  

As the BPT classification is based on narrow-line ratios, we removed the Type 1 AGNs (QSO) from the sample using the spectroscopic classification from the SDSS DR8.We also removed galaxies without SFR estimation available in the MPA-JHU catalogue, yielding a total number of 23816 galaxies contained in the BPT-NII diagram. We do not add the type 1 AGN back to our sample even though their AGN nature is secure because the accretion disk emission may bias their stellar mass estimates.  

Figure \ref{fig2} shows the BPT-NII diagram for our galaxy sample. Galaxies found below the dashed line \citep{2003Kauffmann} are star-forming, between dashed and solid line \citep{2001Kewley, 2006Kewley} are composite, and above the solid line are AGN. Using results of \citep{2007Schawinski}, AGN are further separated between Seyfert 2 (above the dotted line) and LINERs (below). Table \ref{tab1} shows the number of galaxies in each classification. We find that the majority (80\%) of the galaxies in our sample show signatures of AGN activity (Composite, Seyfert 2 or LINER).  

\begin{table*}
\caption{The statistical distribution of galaxies $(\log M_{\star}=10.73- 11.03$ $ \rm{M_{\odot} })$ based on morphological and spectroscopic classification}
 \begin{tabular}{|p{1cm}|cp{1.8cm}|cp{1.5cm}|cp{1.4cm}|cp{1.2cm}|cp{1.2cm}|cp{1cm}|cp{1.5cm}|c|}
\hline 

& & Number & \multicolumn{3}{c}{SFR-stellar mass} &\multicolumn{3}{|c|}{Colour-stellar mass} \\  \cline{4-6} \cline{7-9} 
 Sample& & of sources   & Main   sequence & Starburst &Quiescent & Blue cloud & Green valley &Red sequence \\ 
  
  \hline
 Galaxy    & All& 44092&  12\% & 1\%&  87\%     &  14\%  & 19\%  & 67\%   \\
 %\cline{2-10}
Zoo   & Elliptical& 5322&   1\% & 0\%&  99\%   &   2\%  & 10\%  & 88\%  \\
 %\cline{2-10}
   & Spiral& 16338&   22\% & 1\%&   77\%   &  26\%  & 28\%  & 46\%  \\
 %\cline{2-10}
   & Uncertain& 22432&     7\% & 1\%&   92\%&    8\%  & 14\%  & 78\%   \\
  \hline
  
BPT-  & All& 23816&  21\% & 3\% &  76\% &   24\%  & 27\%  & 49\%  \\
%\cline{2-10}
 NII&star-forming & 4733&     61\% & 11\%  &  28\%&    61\%  & 26\%  & 13\%      \\ %\cline{2-10}
 &Composite & 9458  & 17\% &  0\% & 83\%&    22\%  & 33\%  & 45\%      \\
%\cline{2-10}
 & LINER& 7893&   2\% &  0\%  &  98\%&    4\%  & 20\%  & 76\%      \\
%\cline{2-10}
 &Seyfert 2 & 1732&   18\% &  0\%  & 82\% &  25\%  & 33\%  & 42\%      \\
 \hline
\end{tabular}
\label{tab1}
\end{table*}

\subsection{WISE AGN} \label{wise}
 NASA's  Wide-field Infrared Survey Explorer (WISE; \citealt{2010Wright}) satellite mapped the whole sky in four mid-infrared (MIR) bands centered at 3.4$\mu m$ (W1), 4.6$\mu m$ (W2), 12$\mu m$ (W3) and 22$\mu m$ (W4). We used the AllWISE Data Release\footnote{http://wise2.ipac.caltech.edu/docs/release/allwise/} to check the existence of IR obscured AGN activity in our sources, and validate our optical AGN spectroscopic classification (see Section \ref{BPTsec}). 
  \begin{figure} 
\includegraphics[scale=0.43]{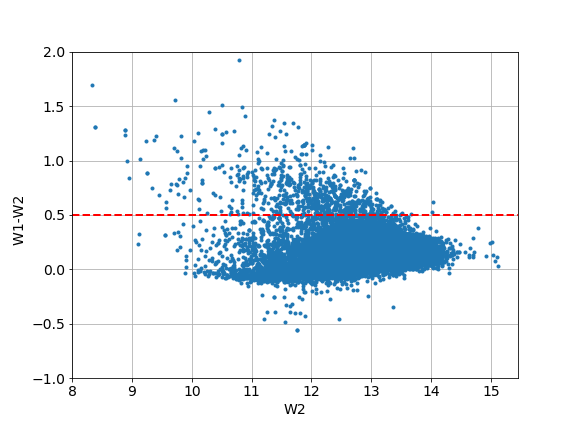}
\caption{(W1-W2) versus W2 colour-magnitude diagram for the mass-selected SDSS sources with a IR counterpart in the WISE catalogue. The adopted colour cut to select AGN sources is shown by the red solid line \citep{2018Assef}.}
\label{fig3}
\end{figure}
We carried out the spectroscopic classification of galaxies, using the BPT diagram, as described in Section \ref{BPTsec}. We then cross-correlated the 24561 galaxies with enough S/N $>$ 3 in their emission lines to be classified in the BPT-NII diagram (See Section \ref{BPTsec} for details) with the WISE source catalogue, by searching for positional matches within 2$''$. We find that 24383 sources (99.3\%) have a WISE IR counterpart. Figure \ref{fig3} shows the (W1-W2) versus W2 colour-magnitude diagram for the selected sources. We classified sources with W1-W2 $>$ 0.5 as AGN following Figure 1 of \citet{2013Assef} and considering the bright W2 magnitudes of our objects. We find that out of the 24383 sources with WISE and SDSS matches, 740 are classified as AGNs using this simple colour selection, 316 of which are spectroscopically classified as Type 1 AGN (QSO). While more strict colour cuts at $(W1-W2)=0.6$ or 0.7 can provide cleaner AGN samples, the selected cut provides a sample with $\sim$ 90\% reliability according to \citet{2018Assef}. 

When applying the BPT classification to the sample of sources identified as AGN through their WISE IR colours, we find that most of them (63\%) are indeed classified as Seyfert 2 and composite (25\%), with a smaller fraction classified as LINER (9\%) and star-forming (3\%) objects. We also find that WISE-selected AGN have significantly larger median SFRs compared to the optically selected AGN. For the WISE-selected AGN we find a median log SFR/[M$_\odot$yr$^{-1}$] of 0.35  (consistent with the results of \citealt{2016Ellison}), while for Seyfert 2 galaxies and LINERs we find -0.24 and -1.14, respectively.

\subsection{Morphological Classification} \label{morp} 
We use the visual morphological classification of galaxies from the Galaxy Zoo\footnote{https://data.galaxyzoo.org/} citizen science project \citep[][]{2008Lintott, 2011Lintott}.  This catalogue provides morphological classification for all galaxies in the SDSS DR7 spectroscopic sample, and is based on a vote fraction threshold of 0.8. Galaxies with lower vote threshold are typically more difficult to classify visually as they tend to have low brightness, low stellar masses, and/or due to the presence of nearby objects (including foreground and background sources) (\citealt{2011Lintott}; \citealt{2009Bamford}). After cross-matching our sample with the Galaxy Zoo catalogue, we find 44092 IOAG candidates (out of 48127 sources), where 12\% are classified as ellipticals, 37\% as spirals, and 51\% are uncertain. All statistics are shown in Table \ref{tab1}.

\section{Analysis and Results}
\subsection{Comparison of morphological and spectroscopic classifications}

\begin{figure*}
\includegraphics[scale=0.49]{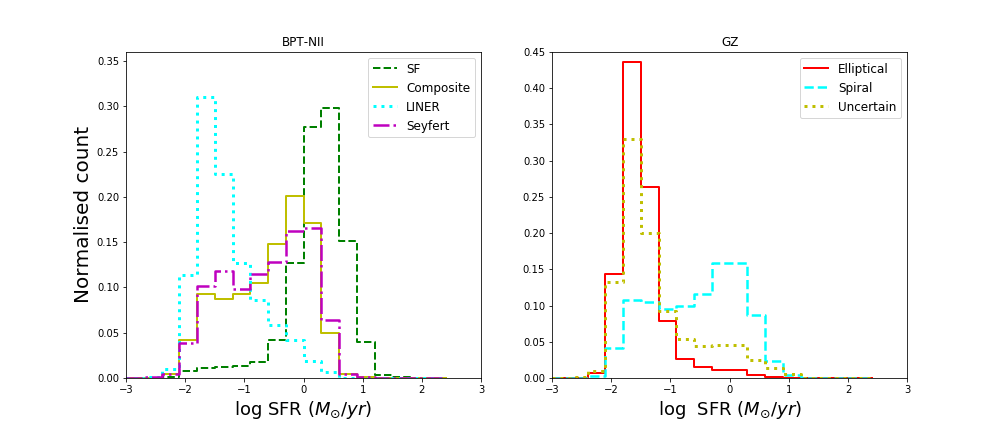} 
\caption{ Distribution of SFR based on the BPT-NII emission-line classification (left-side). The spectroscopic classified galaxies corresponds to SF (blue dot lines), composite (green dot-dashed lines), LINER (red dashed lines) and Seyfert 2 (violet solid line). Distribution of SFR based on the Galaxy Zoo classification (right panel). Elliptical, spiral and uncertain classes are marked with red (solid lines), blue (dashed lines) and green (dot lines), respectively.}
\label{fig4}
\end{figure*}

Figure \ref{fig4} shows the distribution of SFR of different spectroscopic and morphological types. As can be seen, galaxies classified as LINERs appear to have low SFRs, and objects spectroscopically classified as star-forming  galaxies show high SFRs, as expected. Objects classified as Seyfert 2 or composite galaxies have similar bi-modal distributions to those objects morphologically classified as spirals, with a peak at $\log(\rm{SFR})\sim 0$, and a second one at log($\rm{SFR})\sim-1.5$. Note that the distribution of SFRs for the star-forming galaxies extends to higher values compared to that of composite and Seyfert 2.

The right panel of Figure \ref{fig4} also shows that most objects classified as ellipticals are located in the low SFR regime, as expected. The distribution is relatively narrow and peaks at $\log \rm{SFR} (M_{\odot}yr^{-1})\sim-1.6$. Similar SFR values are found for the galaxies with 'uncertain' classification. This likely implies that these objects are either faint ellipticals or S0 galaxies with almost negligible star formation activity. Spiral galaxies are instead found to have a bi-modal distribution, with the bulk of the sources showing relatively high SFRs, namely  log(SFR) $\gtrsim$ 0, and the second group of sources with low SFRs peaking at $\log(\rm{SFR})\sim-1.5$, similar to elliptical and uncertain galaxies, and with no gap between peaks.

With respect to the spectroscopic classification, we find that 64\% (2802) of the star-forming galaxies are classified as spirals and only 3\% (115) as ellipticals. The remaining 33\% (1443) are classified as `uncertain' by Galaxy Zoo. In the case of the composite class, we find that 56\% (4939) are spirals and 8\% (668) ellipticals, whereas, for LINERs, we find that 35\% (2573) of them are classified as spirals and 14\% (1058) as ellipticals. Finally, we find that for Seyfert 2 galaxies,  53\% (857) are spirals and 5\% (81) are ellipticals. In the case of LINERs, the lower number of spirals compared to the other spectroscopic classes tends to agree with previous studies that find that they are mainly hosted by early-type galaxies \citep[][ and references therein]{2008Ho}.  

\subsection{Relation to the star-forming main sequence}
\begin{figure*}
\includegraphics[scale=0.47]{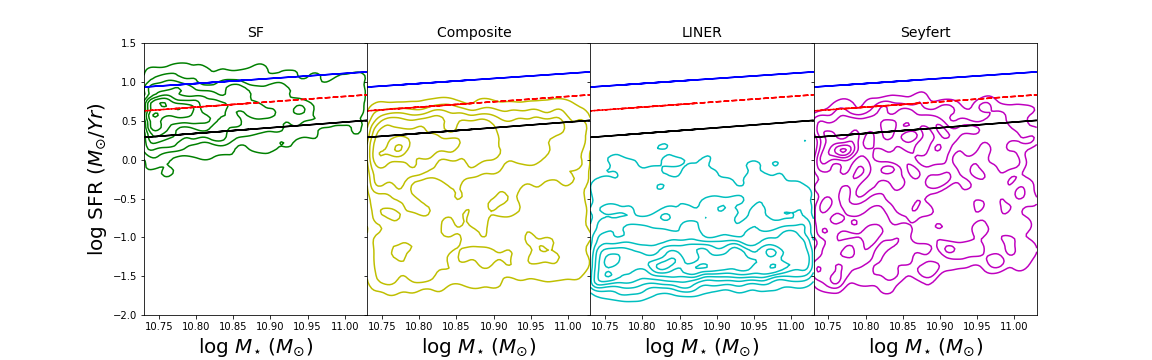} 
\includegraphics[scale=0.47]{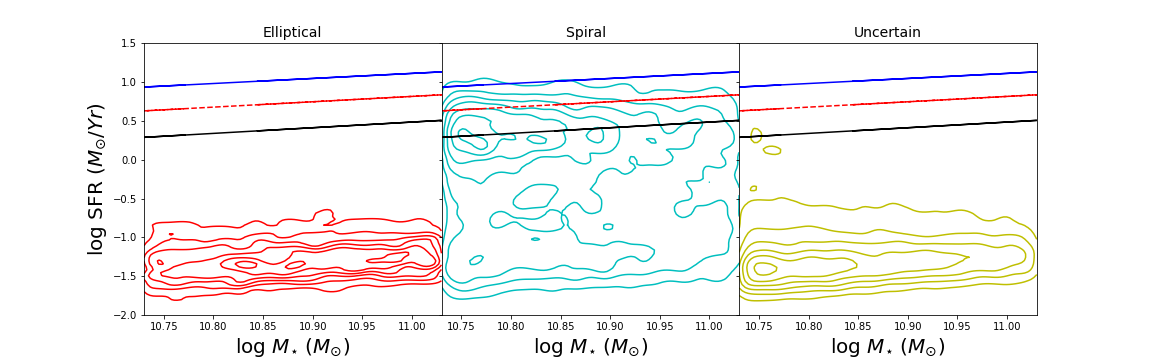} 
\caption{SFR as a function of stellar mass for the IOAG candidates in each spectral class (top panels) and for the morphological classification (bottom panels). A red dashed line in each panel represents the local main-sequence of star formation  from \citet{2012Whitaker}.  The blue and black solid lines indicate the typical scatter around the MS of 0.3 dex.}
\label{fig5}
\end{figure*}

Figure \ref{fig5} shows the SFR versus stellar mass diagram for the sample of IOAG candidates separated both by spectroscopic and morphological class, compared to the main sequence of star formation at $z=0$ \citep{2014Whitaker}. Since, by construction, our sample covers a relatively small stellar mass range, these diagrams show similar distributions as those shown in Figure \ref{fig4}. The main sequence (MS) in this short stellar mass range appears relatively flat, and galaxies with log(SFR) $\lesssim 0.4$ will be located below the lower boundary of the MS. For the width of MS, we used $\pm$ 0.3 dex (black and cyan colour lines in all Figures below), found in different works to be the proper  1$\sigma$ boundaries \citep[][and references therein]{2016povic}. Galaxies spectroscopically classified as star-forming are typically consistent with the MS (61\%), with only 11\% (28\%) being located above (below) the MS. Galaxies spectroscopically classified as composite, LINER or Seyfert 2 have lower SFRs, with only 17\%, 2\%, and 18\%, respectively, being within the bounds of the MS and all the rest below. For the morphological classification, we found that all the ellipticals lie well below the MS, whereas the spiral class shows a bi-modality (already seen in the SFR histograms) with most of them located on the MS and the second cluster of sources well below the MS. Spirals below the MS are likely due to a combination of dust reddening and misclassified S0 galaxies. Yet it is possible that some objects are indeed spiral galaxies effectively in the process of quenching their star-formation (e.g., \citealt{Masters2010}; \citealt{Hao2019}; \citealt{Mahajan2020}).

To visualize more clearly the number of sources of each class above, in, or below the MS, we compute the specific SFR (sSFR $=$ SFR/$M_*$), normalized by the sSFR of the MS at $z=0$ at a given stellar mass. This normalized sSFR ($\delta_{\rm{MS}}=$ sSFR/sSFR$_{\rm MS} (M_{*}))$ represents the distance of a given galaxy (with stellar mass $M_{*}$ and redshift $z$) to the MS, and thus values log($\delta_{\rm{MS}})>0.3$ or $<0.3$ will be associated to starbursts or quiescent galaxies, respectively. 
\begin{figure*}
\includegraphics[scale=0.46]{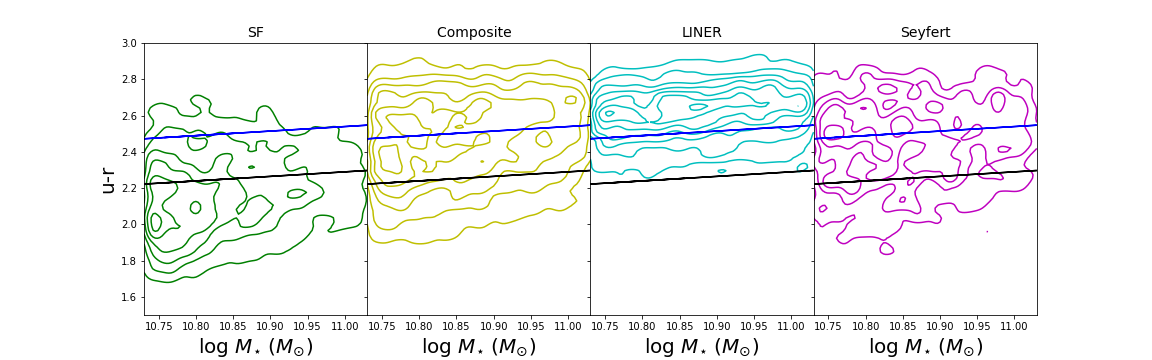}
\includegraphics[scale=0.46]{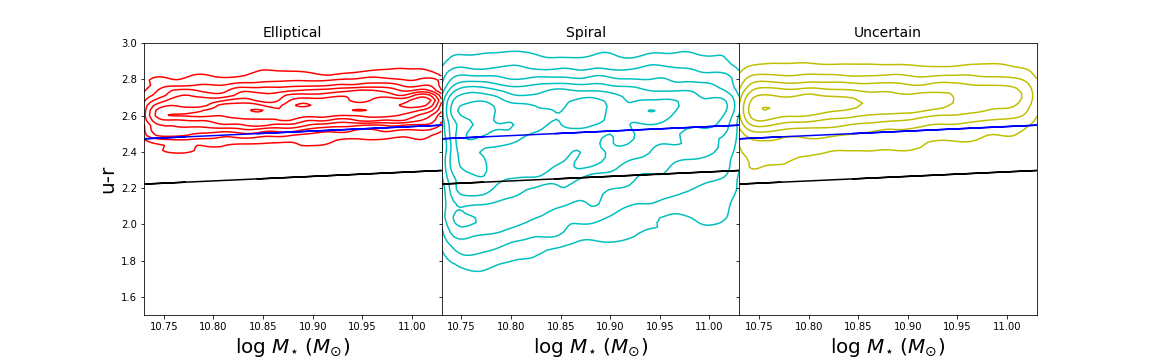}  
\caption{Distribution of IOAG candidates in the rest-frame $u-r$ dust-corrected colour-mass diagram for the spectroscopic (top) and morphological (bottom) classes. The area between the blue and black lines delineate the region occupied by galaxies in the green valley, according to the prescriptions from \citet{2014schawinski}. Red sequence and blue cloud galaxies are located above and below the blue and black solid lines, respectively.}
\label{fig7}
\end{figure*}

We found that from the spectroscopic (morphological) samples of IOAG candidates considered, 76\% (87\%) and 21\% (12\%) are located below and on the MS, respectively, with only 3\% (1\%) of the sources being in the starburst regime.

\subsection{Colour - stellar mass diagram}
In this section, we study the distribution of the rest-frame $u-r$ colour associated with the spectroscopic and morphological classification of the IOAG candidates. We follow \citet{2014schawinski} work and correct the observed optical fluxes for dust reddening using the OSSY Database catalogue\footnote{http://gem.yonsei.ac.kr/~ksoh/wordpress/} and the \citet{1989Cardelli} extinction law. 

Figure \ref{fig7} shows the rest-frame dust-corrected colour-stellar mass diagram for the IOAG candidates split by BPT spectroscopic (top panels) and morphological (bottom panel) class. The solid lines illustrate the location of the ``green valley'', from \citet{2014schawinski}, in the selected stellar mass range using extinction corrected colours (see their Fig. 3). Galaxies located between, below and above these lines are considered to be in the ``green valley'', ``blue cloud'' and ``red-sequence'', respectively.

For star-forming galaxies, we find that 61\%, 26\% and  13\%  of sources are located in the blue cloud, green valley and red sequence, respectively. Composite galaxies are found to be mostly in the red-sequence (45\%), with a significant population located in the green valley (34\%). Similarly, 76\% and 20\% of LINERs are located in the red sequence and green valley, respectively, whereas Seyfert 2 galaxies are found to be relatively uniformly distributed with 42\%, 33\% and 25\% in the red-sequence, green valley and blue cloud, respectively.

For the morphological classification, we found that elliptical galaxies are mostly (88\%) located in the red sequence, as expected, and in the green valley (10\%). For spiral galaxies, 46\% and 28\% are located in the red sequence and green valley respectively. Finally, the uncertain galaxies are mainly located in the red sequence (78\%) and the green valley (14\%).  

\subsection{WISE colour-colour diagram}\label{34}

\begin{figure*}
\includegraphics[scale=0.46]{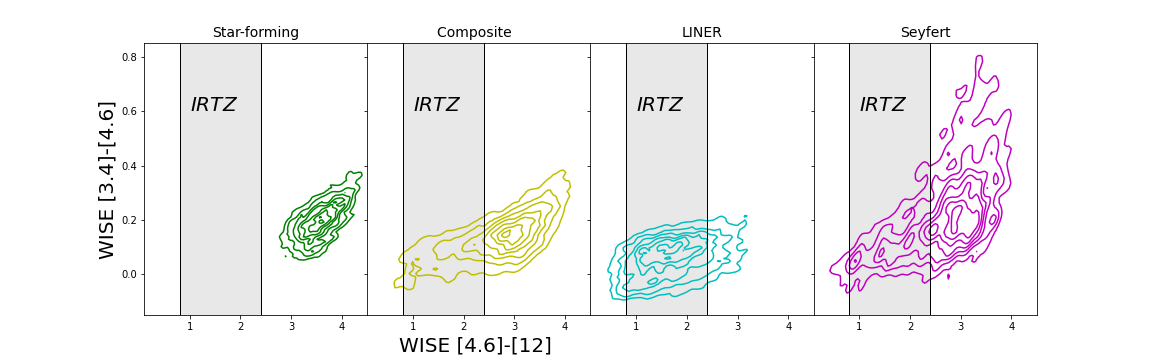}
 \includegraphics[scale=0.46]{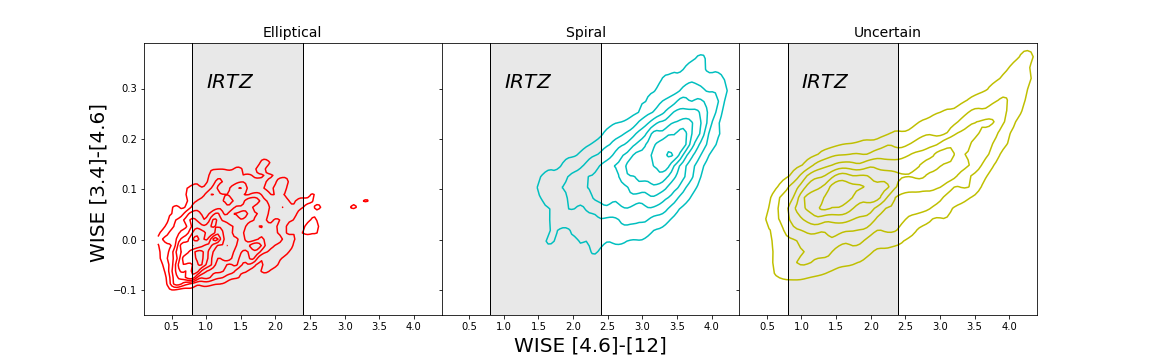}  

\caption{The WISE colour-colour ([4.6]-[12] vs. [3.4]-[4.6]) diagram of spectroscopic (top) and morphological (bottom) classes.  The WISE W2-W3 colour in the range 0.8 - 2.4 is described as an infrared transition zone (IRTZ, \citealt{2014Alatalo}).}
\label{wc}
\end{figure*}

The WISE colour-colour diagram classifies astronomical objects according to their W1-W2 and W2-W3 colours (see  \citealt{2010Wright}). \citet{2014Alatalo} noticed a significant bi-modality in the W2-W3 colour of galaxies that effectively separates early- and late-type objects. They refer to the colour gap between these populations as the Infrared Transition Zone (IRTZ). Remarkably, galaxies within the IRTZ are not necessarily within the optical green valley, since optically selected green valley objects tend to fall closer to star-forming galaxies in the WISE colour-colour diagram \citep{2014Alatalo}. Figure \ref{wc} shows the WISE colour-colour distribution of our sources. We find that 96\% of the star-forming galaxies are located in the red region (i.e., redwards of the IRTZ), and the remaining 4\% are within the IRTZ. We also find that 68\% of composite galaxies lie in the red region, 28\% within the IRTZ and 4\% in the blue region (i.e., bluewards of the IRTZ). Regarding LINERs, 73\% fall within the IRTZ, 20\% in the red region and 7\% in the blue region. Similarly, we find that 70\% of Seyfert 2 galaxies are located in the red region and 28\% in the IRTZ. In terms of the morphological classification, we find that 64\% and 33\% of the elliptical galaxies are located in the IRTZ and in the blue region, respectively, and 66\% and 32\% of the spiral galaxies are located in the red region and in the IRTZ, respectively. We also find that 69\% and 21\% of the uncertain classified galaxies are located in the IRTZ and in the red region, respectively.

\subsection{D$_n$4000 distribution}\label{35}
\begin{figure*}
\includegraphics[scale=0.51]{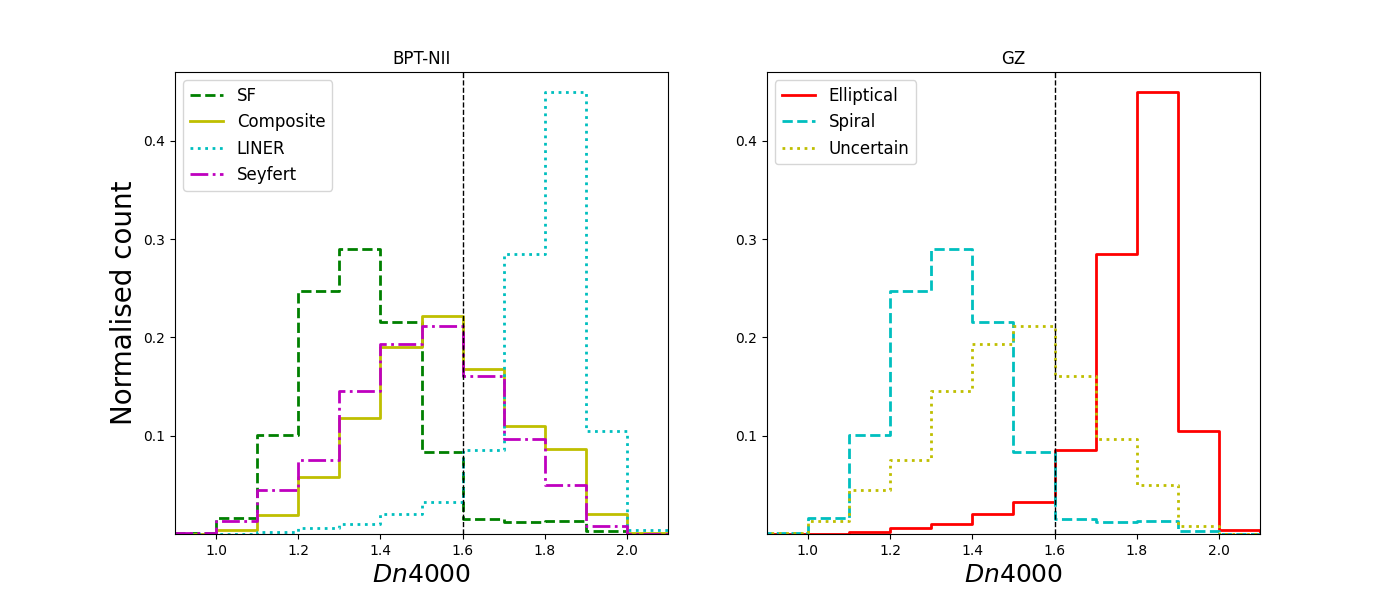} 
\caption{Distribution of D$_n$4000 index for the spectroscopic (left) and morphological (right) classes. The black vertical dashed line shows results of \citet{Li2015} for separating those galaxies taht are centrally star-forming (D$_n$4000 $< $ 1.6) are centrally quiescent (D$_n$4000 $> $ 1.6). }
\label{Dn4}
\end{figure*}

Previous studies have shown a clear connection between the star formation history of galaxies and the stellar absorption line indices, such as D$_n$4000 break and $H_{\delta}$ \citep{2003Kauffmann}. These two indices have been used as indicators of the age of the stellar populations in galaxies.  In addition, D$_n$4000  has been used as an indicator of SFR (\citealt{Brinchmann}; \citealt{2007salim}). In this section we study the distribution of the D$_n$4000 index. Figure \ref{Dn4} shows the distribution of D$_n$4000 in our sample in relation to both, spectroscopic (left) and morphological (right) classifications. Distributions are consistent with results obtained in Figure \ref{fig4}, as expected, taking into account that MPA-JHU SFRs have been measured using the  D$_n$4000 index (\citealt{Brinchmann}).   \citet{Li2015}   used D$_n$4000 to differentiate between galaxies that are centrally star-forming (with D$_n$4000\,$<$\,1.6) or centrally quiescent (with D$_n$4000\,$>$\,1.6). We find more galaxies to be centrally quiescent in both the spectroscopic (62\%) and morphological (72\%) samples. According to this criterion, star-forming galaxies are more likely to be classified as centrally star-forming (87\%), while composites (61\%), Seyfert 2 (58\%), and LINERs (95\%) are more commonly classified as centrally quiescent. Ellipticals (97\%), spirals (62\%), and uncertain (86\%) galaxies in our sample are all found to have predominantly centerally quiescent star-formation activity. 

\begin{figure*}
\includegraphics[scale=0.53]{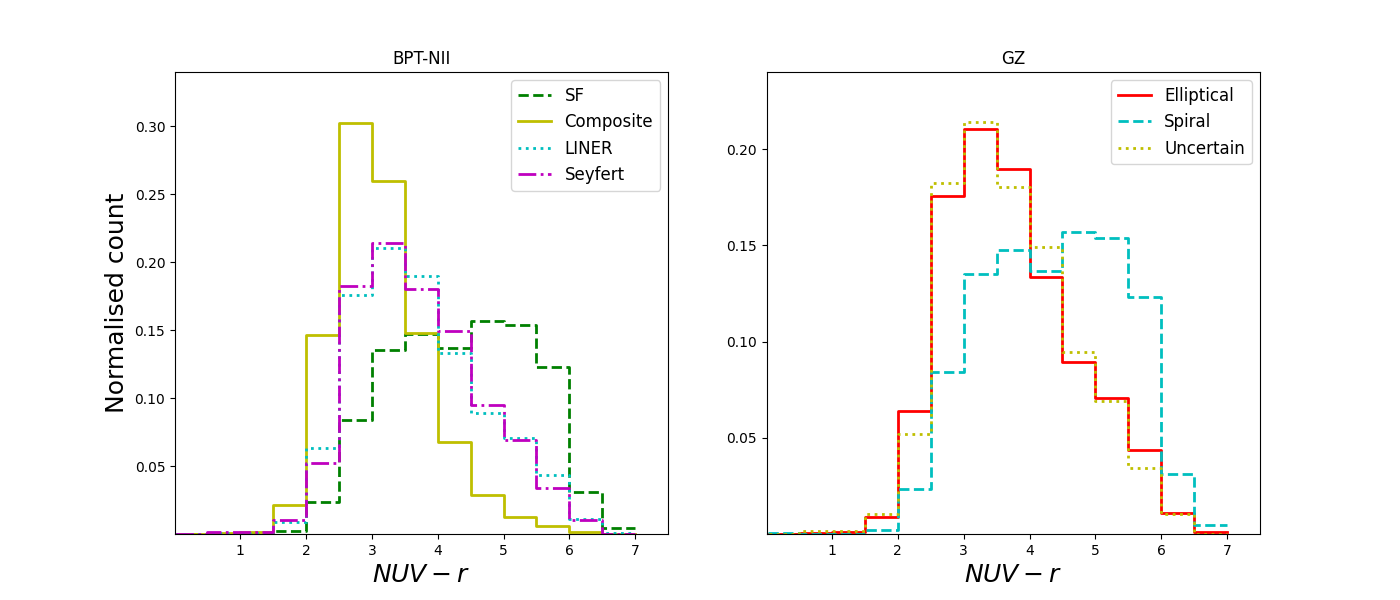} 
  
\caption{Distribution of NUV\,-\,r for of BPT-NII emission-line classification (left-side) and Galaxy Zoo morphological classification (right panel).}
\label{uvd}
\end{figure*}

\subsection{UV colour distribution}
 We use the GALEX GR6+7 \citep{Bianchi_2017} AIS to study the NUV\,-\,r colour distribution of our sample. We find counterparts for 67\% of the spectroscopically and 54\% of the morphologically classified sources. The reason for this is (1) the difference in depths between SDSS and GALEX, and  (2) a large fraction of our sources do not have active star formation or have older stellar populations.  Figure \ref{uvd} shows the normalized  NUV\,-\,r colour distribution for our sample split by classification. The left panel shows that star-forming galaxies have a broader distribution of colours ($NUV\,-\,r$ $\sim3 - 6$ mag), while composite galaxies have a narrower distribution peaked at NUV\,-\,r $\sim2.7$ mag. LINERS and Seyfert 2 galaxies have similar distributions,  peaked at $NUV\,-\,r$ $\sim3.3$ mag. The right panel of Figure \ref{uvd} shows the galaxies split by their morphological classification. Elliptical and uncertain galaxies show similar distributions with peaks at $NUV\,-\,r$ $\sim3.3$ mag, whereas spiral galaxies have broader distribution spanning NUV\,-\,r $\sim3 - 6$ mag.

\section{Discussion}
Observations and simulations suggest that massive galaxies are dominated by spheroidal components and have predominantly old stellar populations particularly in their central regions (e.g., \citealt{Lucia_2012}; \citealt{IbarraMedel2016}; \citealt{James_2016};  \citealt{Morselli2016}; \citealt{Nelson2016}; \citealt{Belfiore2018}; \citealt{AvilaReese2018}; \citealt{Tacchella2018}). \citet{2013Perez} studied the assembly history of galaxies for different spatial regions (nucleus, inner $0.5R_{50} - R_{50}$, and outer region $> R_{50}$).  They find that for the stellar mass range we study here, the great majority of the stars, amounting to 80\% of the stellar mass, at the central region ($<0.5R_{50}$) were assembled a long time ago ($\sim$7.1 Gyr) indicating that low SFRs were in place through the rest of their lifetime, and thus quenching of the star-forming activity must have occurred. The remaining 20\% of the stellar mass would have formed at low SFR and assembled in the past $\sim 2-3$ Gyr. In these massive galaxies, the mean stellar age in the central region is greater than in the outer regions, as discussed earlier (see Section \ref{s1}) , indicating that the bulk of the star-formation activity in the central regions was quenched at earlier times.

In this work, we have analysed the spectroscopic and morphological properties of a sample of SDSS galaxies selected to be in the stellar mass range typical of IOAG candidates \citep{2013Perez}.  
 
We have spectroscopically classified IOAG candidates using the BPT-NII diagram and found that 20\% of them are classified as SF, 40\% as composite, 33\% as LINERs and 7\% as Seyfert 2. These fractions are different from those obtained by  \citet{2016Leslie} using the same data MPA-JHU catalogue, but for the entire SDSS sample. For selecting their sample, these authors used a S/N criteria of H$\alpha$ line > 3 and three different BPT diagrams, although only 17\% of the \citet{2016Leslie} sample are ambiguous galaxies (i.e. do not fall into the same class in all 3 BPT diagrams).  They found for the same redshift range that 60.4\% of galaxies are SF, 12.2\% composite, 6.5\% LINER, and 4.1\% Seyfert 2.  However, regarding the distribution of different spectroscopic types around the MS, we find similar results to \citet{2016Leslie}, where the SF galaxies are mainly located on the MS, composites and Seyferts 2 on and below the MS, and LINERs well below the MS. They also suggested that this sequence supports an evolutionary pathway for galaxies in which star formation quenching by AGN plays a key role. In our study, we found that IOAG candidates with possible presence of AGN activity were significantly below the MS. Specifically, we found that 83\%, 82\%, and 98\% of those classified as composite, Seyfert 2 and LINERs, respectively, were below the MS. We note that out of those galaxies selected as LINERs, up to 79\% could be photoionised by post-AGB stars instead of AGN activity if we use the EW $<$ 3 \AA $ $ criterion proposed by \citet{Fernandes2011} to identify them. If this was to be the case, however, it would not alter the main conclusions of our study.

 Additionally, we found that although a higher fraction of IOAG candidates are spiral galaxies, those that hosted LINERs are most likely ellipticals compared to the other spectral classes, while those classified as Seyfert 2 galaxies where primarily spirals. While the fact that large fractions of IOAG candidates had unclassified Galaxy Zoo morphologies makes these results difficult to interpret, our analysis suggests that IOAG candidates classified as Seyfert 2 are still spiral galaxies, just as those classified as SF, but have quenched SFRs, comparable to those hosting LINERs. This, combined with the fact that AGN activity is more prevalent in IOAG candidates than in galaxies at other stellar masses (as per the comparison above with \citealt{2016Leslie}), suggests that these galaxies could be important to study for improving our understanding of galaxy evolution.

We have also investigated the distribution of ellipticals and spirals in relation to the green valley. We can compare our findings, with the results obtained by \citet{2014schawinski} using the entire SDSS sample at $z < 0.1$. For the same redshift range, we found that galaxies classified as ellipticals in Galaxy Zoo are 88\% in the red sequence, 10\% in the green valley, and 2\% in the blue cloud. \citet[][ see their Table 1]{2014schawinski}, found a slightly bluer distribution for elliptical galaxies, with 82.5\%, 12.4\%, and 5.2\% being in the red sequence, green valley, and blue cloud, respectively. For spirals in our work, we find that 46\% are in the red sequence, 28\% in the green valley, and 26\% in the blue cloud, in contrast to  \citet{2014schawinski} who found 74.1\%, 18.9\%, and 7.0\%, respectively. This result suggests that IOAG candidates are mainly based in the red sequence and the green valley, with a significant fraction being spirals with quenched star-formations. 

 These results are in line with those obtained in Section \ref{34} when studying WISE colour as a significant number of galaxies are located within the IRTZ (64\% of ellipticals, 32\% of spirals, and 69\% of galaxies that remain unclassified morphologically). As concluded by Alatalo et al. (2014), galaxies laying in the IRTZ are mainly in their late stages of transitioning across the optical green valley, shedding the last of their remnant interstellar media. This again is in line with the analysis of D$_n$4000 break in Section \ref{35}, where we found that most of the  galaxies in our sample have centrally quiescent star-formations (62\% and 72\% of all galaxies classified spectroscopically and morphologically, respectively).

Our results are in excellent agreement with the recent study by  \citet{McPartland2018}. They found that the fraction of galaxies hosting AGN activity (composite, LINERs and Seyfert 2) increases towards the most massive end, making up to 60\% of galaxies with $\log M_{\star} >$ 10.5 $M_{\odot}$. We find that 80\% of IOAG candidates with an spectral classification host AGN activity. As explained above, we find consistent evidence that AGN host galaxies are located below the MS and present redder colours , but yet they are more likely to have spiral morphologies.  This could imply that AGN activity in IOAG candidates is associated with galaxies that have recently shut down star-formation activity, but are still in the process of transforming their morphologies from spirals to ellipticals.

\section{Conclusions}

We have presented a study of galaxies with stellar masses in the range $(\log M_{\star} = 10.73 - 11.03 M_{\odot} )$  and $z<0.1$ selected from the SDSS DR8. We use stellar masses and SFRs from the MPA-JHU catalogue, together with morphological classification from Galaxy Zoo. Local galaxies in this stellar mass range were identified by \citet{2013Perez} as having assembly times in their inner regions $(R_{50}<0.5)$ two times shorter than in their outskirts $(R_{50}>0.5)$. We refer to these galaxies as Inside-Out Assembled Galaxies or IOAG candidates. We studied the morphological properties of our galaxies by using the Galaxy Zoo classifications, as well as their spectroscopic properties based on their BPT-[NII] diagram classifications.  We find that:
\begin{itemize}
\item  Using strong enough emission lines with S/N $>$ 3, 20\% of  IOAG candidates are classified as star-forming galaxies, 40\% as composites, 33\% as LINERs, and 7\% as Seyfert 2. This suggests that 80\% of IOAG candidates are not pure star-forming as have been seen previously for the entire SDSS population at $z < 0.1$.
\item IOAG candidates classified as star-forming galaxies have spiral morphologies and are located in the MS of star formation as expected, whereas Seyfert 2 and composites have spiral morphologies, but quiescent SFRs, which may point to the idea that the AGN could be related to their SFR quenching and evolution. In addition, taking into account the high fraction of galaxies with AGN, it might be that AGN activity has an important role in quenching SF in IOAG candidates.
\item We find that IOAG candidates classified as LINERs show the lowest SFRs (median log SFR/[M$_\odot$yr$^{-1}$] of -1.14), and that those with a morphological classification are more commonly spirals (35\%) than ellipticals (14\%).
\item Most of IOAG candidates are spirals rather than ellipticals in the Galaxy Zoo classification scheme, independently of their spectroscopic type (therefore including LINERs). Since they are mainly located in the green valley and the red sequence on the rest-frame colour-stellar mass diagram, with a significant fraction being inside the infrared transitional zone using WISE colours,  we expect these spirals to be the early-type ones. 
 \end{itemize} 
  Our findings suggest that a high fraction of IOAG candidates are transition galaxies. AGN in this stellar mass range have systematically lower star-formation rates than star-forming galaxies, suggesting AGN activity may be related to this quenching. Galaxies at the stellar mass range of $\log M_{\star} = 10.73 - 11.03$ $M_{\odot}$ moving from star-forming to quiescent, and from the blue cloud to the red sequence and/or to recently quenched galaxies.
 
\section*{Acknowledgements}
We thank the anonymous referee for important suggestions and useful comments, which helped us to improve the paper.  We highly appreciate discussions with Prof. Enrique P\'erez that contributed to development of this paper. 

DZ acknowledges support from the European Southern Observatory - Government of Chile Joint Committee through a grant awarded to Universidad Diego Portales.

We thank the Ethiopian Space Science and Technology Institute (ESSTI) under the Ethiopian Ministry of Innovation and Technology (MOIT) for all the financial and technical support. DZ gratefully acknowledge the support from Debre Birhan University. MP in addition acknowledges financial support by the Spanish MEC under grant AYA2016-76682-C3-1-P and financial support from the State Agency for Research of the Spanish MCIU through the "Center of Excellence Severo Ochoa" award to the Instituto de Astrof\'isica de Andaluc\'ia (SEV-2017-0709). M.A. has been supported by the grant “CONICYT+PCI+REDES 190194”. This work was supported by grant "CONICYT+PCI+INSTITUTO MAX PLANCK DE ASTRONOMIA MPG190030". RJA was supported by FONDECYT grant number 1191124.

The data presented in this paper are the result of the efforts of the Galaxy Zoo volunteers, without whom none of this work would be possible. Galaxy Zoo has been supported by  The  Leverhulme Trust.

Funding for SDSS-III has been provided by the Alfred P. Sloan Foundation, the Participating Institutions, the National Science Foundation, and the U.S. Department of Energy Office of Science. The SDSS-III web site is http://www.sdss3.org/.

SDSS-III is managed by the Astrophysical Research Consortium for the Participating Institutions of the SDSS-III Collaboration including the University of Arizona, the Brazilian Participation Group, Brookhaven National Laboratory, Carnegie Mellon University, University of Florida, the French Participation Group, the German Participation Group, Harvard University, the Instituto de Astrofisica de Canarias, the Michigan State/Notre Dame/JINA Participation Group, Johns Hopkins University, Lawrence Berkeley National Laboratory, Max Planck Institute for Astrophysics, Max Planck Institute for Extraterrestrial Physics, New Mexico State University, New York University, Ohio State University, Pennsylvania State University, University of Portsmouth, Princeton University, the Spanish Participation Group, University of Tokyo, University of Utah, Vanderbilt University, University of Virginia, University of Washington, and Yale University.

This publication makes use of data products from the Wide-field Infrared Survey Explorer, which is a joint project of the University of California, Los Angeles, and the Jet Propulsion Laboratory/California Institute of Technology, funded by the National Aeronautics and Space Administration.

\section{Data Availability}
 No new data were generated or analysed in support of this research. The data used in this article are available from the public sources mentioned in the article (or references therein).
 
%The Acknowledgements section is not numbered. Here you can thank helpful colleagues, acknowledge funding agencies, telescopes and facilities used etc. Try to keep it short.

%%%%%%%%%%%%%%%%%%%%%%%%%%%%%%%%%%%%%%%%%%%%%%%%%%

%%%%%%%%%%%%%%%%%%%% REFERENCES %%%%%%%%%%%%%%%%%%

% The best way to enter references is to use BibTeX:
\bibliographystyle{mnras}
\bibliography{ref} % if your bibtex file is called example.bib

% Alternatively you could enter them by hand, like this:
% This method is tedious and prone to error if you have lots of references
%\begin{thebibliography}{99}
%
%\bibitem[\protect\citeauthoryear{Author}{2012}]{Author2012}
%
%
%
%
%\bibitem[\protect\citeauthoryear{Author}{2012}]{Author2012}
%Author A.~N., 2013, Journal of Improbable Astronomy, 1, 1
%\bibitem[\protect\citeauthoryear{Others}{2013}]{Others2013}
%Others S., 2012, Journal of Interesting Stuff, 17, 198
% 
%\end{thebibliography}

%%%%%%%%%%%%%%%%%%%%%%%%%%%%%%%%%%%%%%%%%%%%%%%%%%

%%%%%%%%%%%%%%%%% APPENDICES %%%%%%%%%%%%%%%%%%%%%

\appendix

%\section{Some extra material}

% Don't change these lines
\bsp	% typesetting comment
\label{lastpage}
\end{document}